\documentclass[twocolumn,showpacs,preprintnumbers,amsmath,amssymb]{revtex4}
\usepackage{graphicx}
\usepackage{graphics}
\usepackage{dcolumn}
\usepackage{bm}
\begin{document}

\title{Quasi-Linear Soft Tissue Models Revisited}
\author{J. S. Espinoza Ortiz, Gilson A. Giraldi, E.A. de Souza Neto, Raul A.
Feij\'{o}o}
%\date{\today}

\begin{abstract}
Incompressibility, nonlinear deformation under stress and
viscoelasticity are the fingerprint of soft tissue mechanical
behavior. In order to model soft tissues appropriately, we must
pursue to complete these requirements. In this work we revisited
different  soft tissue quasi-linear model possibilities in trying
to achieve for this commitment.
\end{abstract}

\affiliation{National Laboratory of Scientific Computing\\
\mbox{Ave.Get\'ulio Vargas 333, 256251-075, Petr\'opolis, RJ,
Brazil}\\ \mbox{jsespino,gilson@lncc.br}}
\maketitle

\section{Introduction}
Different mechanical properties of biological tissues depend on
mineral content. Thus, roughly, it can be distinguished two
classes of biological tissues. Bones and tooth content mineral,
they conform the group known as hard tissues. Whereas, skin,
muscle, blood vessel and lungs conform the second group called of
soft tissues. They do not content mineral, so they are much
deformable than hard tissues.

\begin{figure}[t]
\centering \vspace*{1.cm}\setlength{\abovecaptionskip}{-3.5cm}
\includegraphics[width=8.cm,height=8.cm]
{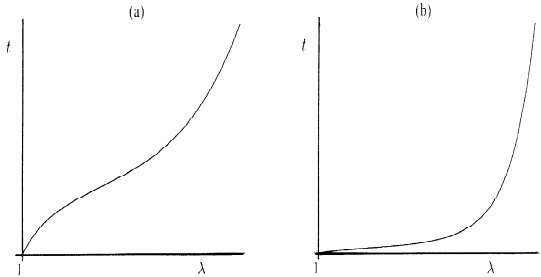} \caption{Typical simple tension response of (a) rubber
and (b) soft tissue. Nominal stress $t>0$ plotted against stretch
$\lambda \ge \,1$\thinspace . } \label{fig:one}
\end{figure}

Modelling the mechanical behavior of soft tissues has much in
common with those techniques used to model rubber elasticity.
Thus, finite deformation theories useful for rubber elasticity are
often used to describe soft tissue mechanical behavior. However,
there are significant differences in the material structure of
soft tissues and rubber elasticity. Moreover, in the way they
respond under applied stress. Soft tissue material achieve
initially large stretch under relatively low level of stress and
subsequent stiffening at higher level extensions. This is shown in
Figure(~\ref{fig:one}), where are compared the typical simple
tension stress-stretch response of rubber (left-hand figure) with
that of soft tissue (right-hand figure). For other side, collagen
fibres distribution leads to pronounced anisotropy in soft tissue,
in contrast with from typical isotropic rubber
\cite{ogden,holzapfel,holzapfel2}.

In biomechanics an specimen (animal) is regarded as an assemble of
rigid bodies connected by joins and muscles. Under this framework,
Newton's law balance of forces maybe could not be efficient to
write equations of motion for each connectors, known their
mechanical properties. Naturally, because the number of particles
to be considered is excessively large, and the way to specify the
interaction forces between particles becomes very complex.
Instead, it is much better to consider the body as a continuum,
searching for a simplified way to specify interaction forces
between particles by means of constitutive equations, and so thus
specifying realistic properties of materials. A stress-strain
relationship describes the mechanical property of a material and
it is therefore a constitutive equation.

Research on modelling soft tissue mechanical behavior has a
growing demand for applications in surgical simulations, pursuing
for in real time fast and precise calculations of tissue
deformations. In trying for these commitments, it has been
introduced models accounting for the continuous nature of soft
tissue. Within the limits of the employed constitutive model the
finite element methods allows for physically correct simulation of
the tissue mechanics \cite {bro-nielsen,keeve,koch}. For other
side, alternative discrete approaches based on spring-mass model
have also been applied\cite{mosegaard,tescher}. It seems that
methods based on the continuum approach are more appropriated to
describe much better realistic soft tissue mechanical behavior,
whereas spring-mass based models being economics are less accurate
than the former. The main issue of our research look for an
appropriate way to establish an accurate comparison between these
methodologies.

In this work we review some physic models applicable to modelling
soft tissue mechanical behavior and we also discuss some
perspective in the field. This work is organized as follow: In
section~\ref{sec:continuum} we present the continuum approach
formalism, and we also consider particular mathematical results
for linear elastic solids and fluids. Important testing methods on
physics necessaries to characterize soft tissue mechanical
behavior, i.e, strain, creep and relaxation, are presented in
section~\ref{sec:testmeth}. Next, in section~\ref{sec:tissueprop}
are reviewed more general soft tissue mechanical properties.
Section~\ref{sec:tissuemath} is devoted to show basic mathematic
models describing nonlinear elasticity and quasi-linear
viscoelasticity, which are related to soft tissue mechanical
behavior. In section~\ref{sec:vibrsyt}, it is presented the basic
vibrational spring-mass model, which is also extended to consider
viscoelastic mechanical behavior. In section~\ref{sec:conclus} we
presented some discussions and perspectives.
\section{\label{sec:continuum}Continuum Approach}
The equations of motion for a continuum were derived by Euler,
using Newton's laws and supported on the follow axioms:
\textit{i}) The material particles form a one-to-one isomorphism
with real numbers in a $3\mathcal{D}$ Euclidean space.
\textit{ii}) The mass distribution is characterized by the density
$\rho $ (the mass per unit volume), it is defined as a piecewise
continuous function over the volume of the continuum.
\textit{iii}) Solely neighbors particle interactions are regarded.
A surface is conformed by particles, it could be considered
arbitrarily infinitesimal. In a continuum, oriented surfaces
interaction can be expressed as a surface traction ( Force per
unit area), that can be computed from a well defined stress tensor
($\sigma _{ij}\,$).

Let us $x_{1},x_{2},x_{3}$ be an inertial cartesian frame of
reference, and an infinitesimal volume by $dx_{1}dx_{2}dx_{3}$.
Let $\sigma _{ij},\,e_{ij}$
be the stress and strain tensor, respectively. Let us denote: $%
v_{i},\,D\,v_{i}/Dt$ and $X_{i}$; the velocity vector, the
acceleration vector and the body force per unit volume,
respectively. We now recall Newton's balance of forces, stating
that the material rate of change of the linear momentum of a body
must be equal to the resultant of forces applied to the body. In
follow, we write it in the form of Euler's equation, for more
details see \cite{fung}:

\begin{eqnarray}
v_{i}\left( \frac{\partial {\rho }}{\partial {t}}+\frac{\partial
{\rho }v_{j}}{\partial {x}_{j}}\right) +\rho \frac{Dv_{i}}{Dt}
=\frac{\partial \sigma _{ij}}{\partial
{x}_{j}}+X_{i}\,.\label{eqn:one}
\end{eqnarray}

Using tensor notation, a repetition of an index in any given term means a
summation over the range of that index,
\begin{displaymath}
\frac{\partial \sigma _{ij}}{\partial {x}_{j}}=\frac{\partial
\sigma _{i1}}{\partial
{x}_{1}}+\frac{\partial\sigma_{i2}}{\partial {x}_{2}}+\frac{
\partial \sigma _{i3}}{\partial {x}_{3}}\,.
\end{displaymath}
In the spatial description,
\begin{displaymath}
\frac{Dv_{i}}{Dt}=\frac{\partial \,v_{i}}{\partial \,t}+v_{j}\frac{\partial
\,v_{i}}{\partial \,x_{j}}\,.
\end{displaymath}

By taking into account the conservation of mass, expressed by the equation
of continuity:
\begin{eqnarray}
\frac{\partial\rho}{\partial{t}}+\frac{\partial\rho\,v_{j}}{\partial{x_{j}}}\equiv\,
\frac{D\rho}{Dt}+\rho\frac{\partial{v_{j}}}{\partial{x_{j}}}\,=0\,,\label{eqn:two}
\end{eqnarray}
Eqn.(~\ref{eqn:one})  becomes,
\begin{eqnarray}
\rho\frac{Dv_{i}}{Dt}=\frac{\partial \sigma _{ij}}{\partial
\,x_{j}}+X_{i}\,
\hspace{1cm}\,\left(i=1,2,3\right)\,.\label{eqn:three}
\end{eqnarray}
This is valid for any continuum, whether it is a fluid or a solid.

In particular, if the material is considered to be incompressible, regarded $%
\rho $ is a constant; we get from the Eqn.(~\ref{eqn:two}),
\begin{eqnarray}
\partial \,v_{j}/\partial \,x_{j}=0\,.\label{eqn:four}
\end{eqnarray}
From here, further development requires specification of the
properties of the material in the form of constitutive equations,
relating stress with strain or strain rate; or strain history. In
order to have a complete theory, all these constitutive
relationship must be incorporated into Eqn.(~\ref{eqn:three}).

Using our faculty perceptions we can verify: booth skin and muscle
have a definite shape, they even can sustain a shear force, and
maintains a quasi-static state (i.e. stress is a function of
strain). Therefore, their mechanical behavior is analogous to that
one of a solid. Whereas considering blood (at least in normal
conditions) we note: it has no shape, it can not sustain a shear
force (because it is in motion), and given its dynamic state; the
stress is a function of the rate of strain. Thus it behaves
mechanically like a fluid.

Next, we derive particular equations for isotropic solid materials
which follow Hook's elastic law and also for Newtonian fluids
using Euler's formalism, incorporating specific constitutive
relationships.

\subsection{Isotropic Hookean Solid}
These kind of materials obey the following stress-strain linear
relationship:
\begin{eqnarray}
\sigma_{ij}=\lambda\,e_{kk}\delta_{ij}+2\,G\,e_{ij}\,;\label{eqn:five}
\end{eqnarray}
which alternative inverse relationship can be re-written as,
\begin{displaymath}
e_{ij}=\frac{1+\nu}{E}\sigma_{ij}-\frac{\nu}{E}\sigma_{kk}\delta_{ij}\,,
\end{displaymath}
where $\sigma_{ij}$ and $e_{ij}$ are the stress and strain tensors, $\lambda$
is the Lam\'e constant, $G$ is the shear modulus, $E$ is the Young's modulus
and $\nu$ is the Poisson's ratio. $\delta_{ij}$ is the Kronecker delta
(unity if $i=j$, otherwise is zero ). We see here that, the strain is being
referred to a configuration of the body in which it is null provided that
the stress is zero.

Substituting Eqn.(~\ref{eqn:five}) into Eqn.(~\ref{eqn:three}) we
arrived to,
\begin{eqnarray}
\rho \frac{Dv_{i}}{Dt}=\lambda \frac{\partial {e}_{\alpha \alpha
}}{\partial {x}_{i}}+2G\frac{\partial {e}_{ij}}{\partial
{x}_{j}}+X_{i}\,.\label{eqn:six}
\end{eqnarray}
Lets now denote the vector displacement of a point in the body by
$\mathit{u}_{i}$, which is measured with respect to an inertial
cartesian frame of reference. If
$\mathit{u}_{i}(x_{1},x_{2},x_{3},t)$ is infinitesimal, we can
define,
\begin{eqnarray}
e_{ij}=\frac{1}{2}\left( \frac{\partial \mathit{u}_{i}}{\partial
\,x_{j}}+\frac{\partial \mathit{u}_{j}}{\partial\,x_{i}}\right)
\,.\label{eqn:seven}
\end{eqnarray}
To the same order of approximation,
\begin{eqnarray}
v_{i}=\frac{\partial \mathit{u}_{i}}{\partial
\,t}\,,\hspace{1cm}\,\frac{Dv_{i}}{Dt}=\frac{\partial^{2}\mathit{u}_{i}}{\partial
\,t^{2}}\,.\label{eqn:eight}
\end{eqnarray}
On substituting all these equations into Eqn.(~\ref{eqn:six}), it
is obtained the well known Navier's equation:
\begin{eqnarray}
\rho\frac{\partial ^{2}\mathit{u}_{i}}
{\partial{t}^{2}}\,=\,G\frac{\partial^{2}\mathit{u}_{i}}
{\partial{x}_{j}\partial{x}_{j}}+\frac{G}{1-2\nu}
\frac{\partial}{\partial{x}_{i}}\frac
{\partial\mathit{u}_{j}}{\partial{x}_{j}}+X_{i}\,,\label{eqn:nine}
\end{eqnarray}
where the Poissons' ratio $\nu =\frac{\lambda }{2\left( \lambda
\,+\,G\right) }$. In this way, we arrived to the basic field
equation of the linearized theory of elasticity. However, living
bodies often take on finite deformations, obey so more complex
nonlinear constitutive equations than the above presented.

\subsection{Newtonian Fluid }
These fluids obey the following stress-strain-rate relationship,
\begin{eqnarray}
\sigma _{ij}=-\mathit{p}\,\delta _{ij}\,+\,
\mu\left(\frac{\partial{u}_{i}}{\partial{x}_{j}}+\frac{\partial{u}_{j}}
{\partial{x}_{i}}\right)\,+\, \lambda
\left(\frac{\partial{u}_{k}}{\partial{x}_{k}}\right)\,\delta_{ij}\,,\label{eqn:ten}
\end{eqnarray}
where $\mathit{p}$ is the pressure, $\mu$ is the dynamic viscosity
( a constant parameter), and $\lambda $ is the bulk elasticity (
second coefficient of viscosity). If the fluid is incompressible,
according to Eqn.(~\ref{eqn:four}), this expression simplified to,
\begin{eqnarray}
\sigma _{ij}=-\mathit{p}\,\delta
_{ij}+\mu\,\left(\frac{\partial{u}_{i}}{\partial{x}_{j}}+\frac{\partial{u}_{j}}{\partial{x}_{i}}\right)
\,.\label{eqn:eleven}
\end{eqnarray}
By substituting it into Eqn.(~\ref{eqn:two}), we obtain the
Navier-Stokes equation, valid for Newtonian fluids,
\begin{eqnarray}
\frac{\partial \mathit{u}_{i}}{\partial{t}}+\mathit{u}_{j}
\frac{\partial\mathit{u}_{i}}{\partial{x}_{j}}=-\frac{1}{\rho}\frac{\partial\mathit{p}}
{\partial{x}_{i}}\,+\,\nu\frac{\partial^{2}\mathit{u}_{i}}
{\partial{x}_{j}\partial{x}_{j}}\,+\,X_{i}\,,\label{eqn:twelve}
\end{eqnarray}
where $\nu\,\left(=\mu /\rho\right)$ is the fluid kinematic
viscosity. Appropriate boundary conditions are necessary in order
to solve the current equation.

Air and water are Newtonian fluids. Non-Newtonian fluids do not
obey constitutive equations (~\ref{eqn:twelve}). Like blood, most
body fluids are non-Newtonian. However at a shear strain rate
above $100\, sec^{-1}$ blood is almost Newtonian.

In general, soft tissues present unique mechanical properties. In the next
section we present some basic test methods useful to describe quantitative
soft tissue mechanical properties.

\section{\label{sec:testmeth}Test Methods}
The tensile test method consists to measure deformation of an
specimen while a force being applied to a sample is gradually
increased. During the stretching process are measured both, the
stress and the strain. The former is obtained taking the ratio of
the force and the material cross sectional area, whereas the
latter is derived dividing the displacement and the original
material length. Analyzing the stress-strain curve provides useful
information: as said the Young's module, the yield point and the
ultimate tensile strength. The Young's module is obtained taking
the slope of the linear stress-strain curve. The yield strength
corresponds to the value of the stress at the yield point,
generically calculated by plotting Young's module at a specified
percent of offset ($0.2\%$). The ultimate tensile strength
corresponds to the highest value of stress on the stress-strain
curve. Computing the area under the stress-strain curve we
determine the fracture energy.

Meanwhile, viscoelastic properties of soft tissues can be
determined by creep and relaxation tests. Creep properties are
determined subjecting the specimen under prolonged constant
tension or compression, and the deformation is measured at
specified times intervals. Then, a creep (deformation) versus time
diagram is plotted, and the creep-rate is obtained by taking the
slope of curve at any point. Provided failure occurs, the test is
finished and the time for rupture is recorded. If the specimen
does not fracture within the test period, creep recovery may be
measured too.

The stress-relaxation of a material is obtained taking the
specimen deformed a given amount and recording the actual stress
over prolonged period of exposure, and the stress-relaxation-rate
is the slope of the curve at any point.
\begin{figure}[b]
\centering \vspace*{.5cm}\setlength{\abovecaptionskip}{-3.cm} %
\includegraphics[width=8.5cm,height=8.5cm]{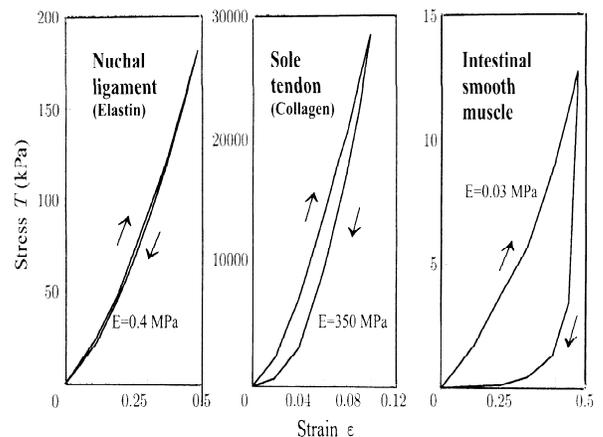}
\caption{Elastin-rich canine nuchal ligament, collagen-rich sole
tendon, and intestinal smooth muscle tensile properties. The sole
tendon shows to be harder than the other two kinds of tissues,
smaller the deformation biggest is the strain. In contrast, the
intestinal smooth muscle shows to be more softness, strain value
of $0.5$ is achieved with more facility than for nuchal ligament
(elastin), it is also more viscoelastic than the others tissues,
because it shows a wider hysteresis loop. } \label{fig:two}
\end{figure}
\section{\label{sec:tissueprop}Soft Tissue Basic Properties}
Once the biologic specimen is subjected to test methods as
described above, their mechanical properties are obtained. In
follow we describe more basic soft tissue mechanical properties.

\subsection{Inhomogeneous Structure}
\begin{figure}[b]
\centering \vspace*{2.cm}\setlength{\abovecaptionskip}{-3cm} %
\includegraphics[width=8.5cm,height=8.5cm]{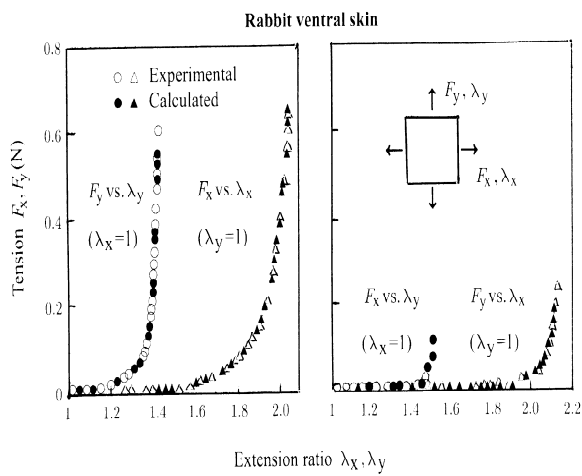}
\caption{Tension-extension ratio relations of rabbit ventral skin.
Different predetermined orthogonal directions are shown in the
inset.\label{fig:three}}
\end{figure}
Biological soft tissues are mainly composed of cells and
intercellular substances. The later consisting of connective
tissues such as collagen, elastin and ground substance(hydrophilic
gel). These components present different physical and chemical
properties, and their contents differ from tissue to tissue and
even from location to location within a tissue. Thus, mechanical
properties depend both on tissue and on site.

Collagen is a protein which gives mechanical integrity and
strength to our bodies, it is presented in different structural
forms in different tissues and organs.

Elastin is also a protein, unlike to collagen has a much less integrated
structure. It has much less strength and much more flexibility than collagen.

Figure (~\ref{fig:two}) shows the stress-strain relationship for
soft tissues rich in collagen, elastin and smooth muscle
(cell)\cite{hasegawa}. There we can see that, the elastin-rich
nuchal ligament has much less strength and much more flexibility
than the collagen-rich sole tendon, whereas the intestinal smooth
muscle is much softer than the other soft tissues, which also
shows a wider hysteresis loop; indicating that it is viscoelastic.

\subsection{Nonlinear Large Deformation and Anisotropy}

Biological soft tissues are mechanically nonlinear, this is enhanced by
assembly of structural component into a tissue. The behavior of arterial
walls at low tension is similar to that of elastin, while the behavior at
high tension is the same as that of collagen.

Collagen and elastin are long-chained polymers, so they are
intrinsically anisotropic. In order to work effectively, not only
their fibers but also cells are oriented in tissues and organs.
Skin for example has very different properties in two orthogonal
directions. This is shown in Figure (~\ref{fig:three}), where
extensions under different applied values of tension were
measured. In this test method, Tension consists to subject the
specimen to pre-determined value of force, i.e. $F_{x,y}$; the
subindex recall for the direction, whereas the extension
$\lambda_{x,y}$ is measured, taking the ratio of material length
under the actual tension and its original one (when the tension is
null). Note here that the sample achieves it original length when
the tension value is zero.

\begin{figure}[t]
\centering \vspace*{-0.5cm}\setlength{\abovecaptionskip}{-1.5cm}
\includegraphics[width=9.cm,height=9.cm]{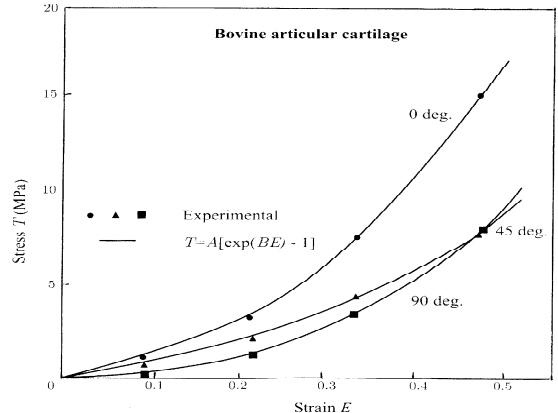}
\caption{Tensile properties of bovine articular cartilage for
different orientations. Here, we can also note certain nonlinear
strain response of the soft tissue under stress.} \label{fig:four}
\end{figure}

Analogously, collagen fibers in the articular cartilage also shows
anisotropic mechanical behavior\cite{tong}. This is presented in
Figure (~\ref{fig:four}), where are shown different stress-strain
curve responses for different orientations, i.e.$0,45,90$ degrees.

\subsection{\label{subsec:visco}Viscoelasticity}
\begin{figure}[t]
\centering \vspace*{2.5cm}\setlength{\abovecaptionskip}{-3.5cm}
\includegraphics[width=9.cm,height=9.cm]{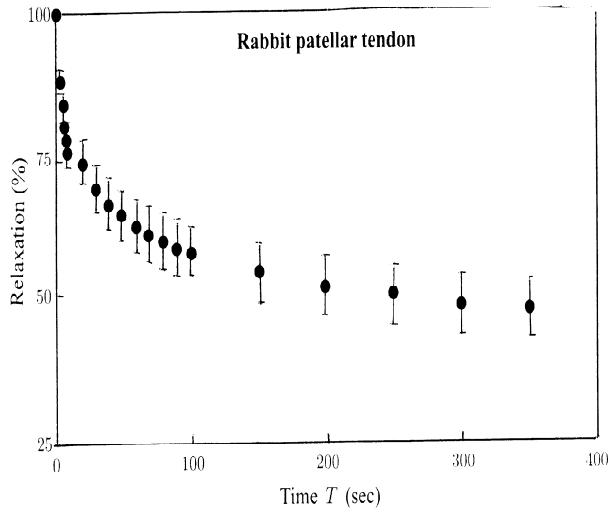}
\caption{Relaxation behavior of rabbit patellar tendon}
\label{fig:five}
\end{figure}
Most rheologist would restrict using of the label non-Newtonian
fluid to substances which, because they possess long structural
relaxation times, are non linear and tend to remember their past
history. For almost all the biological soft tissues the curve
stress-strain exhibit hysteresis loop, as it was shown in Figure
(~\ref{fig:one}) for elastin, collagen and soft muscle, meaning
that they are viscoelastic. However, viscoelastic behavior is
distinguished more clearly by relaxation test. For example, as it
is shown in Figure (~\ref{fig:five}), when the patellar tendon is
elongated a given amount and the actual stress does not stay at a
specific level but decreases rather rapidly at first and then
gradually\cite{yamamoto}.

Generically speaking, under different strain rates (test speed)
biological soft tissues are mechanically not very sensitive. In
addition, the area of the hysteresis loop does not depend on the
strain rate.

\subsection{Incompressibility}
Most biological tissues hardly change their volume even if load is
applied, so they are almost incompressible\cite{chuong}. They have
a water content of more than 70 $\%$. However, articular cartilage
is an exception, because is a micro-porous tissue, therefore water
can enter an leave pores depending upon load\cite{woo}.

The incompressibility assumption is a very important ingredient in
constitutive laws formulation, imposing that adding overall
principal strains is always zero.

\section{\label{sec:tissuemath}Mathematical Description of Mechanical Behavior}
\subsection{\label{tissuemath-spring}Tensile Behavior}
\begin{figure}[b]
\centering \vspace*{2.5cm}\setlength{\abovecaptionskip}{-3.cm} %
\includegraphics[width=9.cm,height=9.cm]{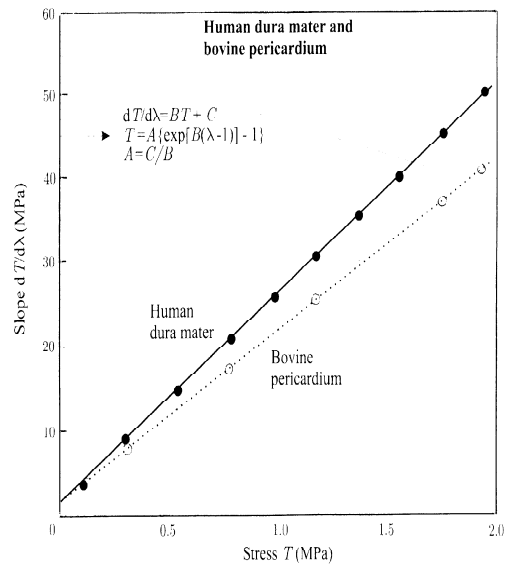}
\caption{Slope of stress-extension ratio curve versus stress of
human dura matter and bovine pericardium} \label{fig:six}
\end{figure}
\begin{figure}[t]
\centering \vspace*{-0.cm}\setlength{\abovecaptionskip}{0.cm} %
\includegraphics[width=8.cm,height=8.cm]{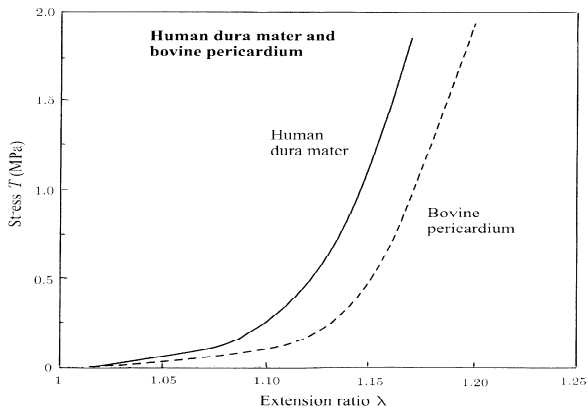}
\caption{Stress-extension curves of human dura matter and bovine
pericardium\thinspace .} \label{fig:seven}
\end{figure}

Biological soft tissue can support large deformations which are
nonlinear and extensive. In reducing the experimental data to a
stress-strain relationship, it is used the nominal stress
$\mathit{T}$ (getting dividing the force by the original cross
sectional area specimen at zero stress). The most striking feature
of this relationship is revealed in Figure (~\ref{fig:six}) when
the slope of the stress-extension curve is plotted against
$\mathit{T}$. We can fit the experimental data to a straight line,
as first approximation,
\begin{eqnarray}
\frac{d\mathit{T}}{d\lambda
}\,=\,B\,\mathit{T}\,+\,C\,,\label{eqn:thirteen}
\end{eqnarray}
where $\lambda $ is the extension ratio, $B$ and $C$ are constants
depending of the material. Prescribed $\mathit{T}=0$, $\lambda
=1$, from Eqn. (~\ref{eqn:thirteen}) we obtain,
\begin{eqnarray}
\mathit{T}\,=\,\frac{C}{B}\left\{\exp{\left[
B\left(\lambda-1\right)\right]}\,-\,1\right\} \,.
\end{eqnarray}
Several other types of soft tissue such as the skin, the muscles,
the ureter, the lung tissue, and articular cartilage are found to
follow similar relationship, Figure (~\ref{fig:seven}). So, it
appears that the exponential type of stress-strain relationship is
common to biological tissue.

Realistically, in vivo tissues are not expose to such a pure tensile force
conditions. Therefore, under more realistic mechanical conditions
multi-axial test are often used to determine soft tissue mechanical
behavior. So in several cases we need more equations. Naturally, for a
three-dimensional organ we need a three-dimensional stress-strain law.

\subsection{Multi-axial Behavior}
No general constitutive equations has been identified for living
tissues. Strain energy functions are used to describe the
mechanical tissue properties. The main idea is that if a
pseudo-elastic strain energy function exist, then the
stress-strain relationship can be obtained by a differentiation.
Let $\mathit{W}$ be the strain energy per unit mass of a tissue,
and $\rho_{0}$ be the density in the zero-stress state. Thus
$\rho_{0}\mathit{W}$ is the strain energy per unit volume of a
tissue in the zero-stress state, this is called the strain-energy
density function. The strain energy $\mathit{W}$ can be expressed
in terms of the components $\mathit{E}_{ij}$ of the Green strain.
In a uni-axial tensile test, the Green strain $\mathit{E}$ in the
tensile direction is given by
\begin{eqnarray}
\mathit{E}=E_{11}=\frac{1}{2}\left(\lambda^{2}-1\right)\,.
\end{eqnarray}
The components $\mathit{S}_{ij}$ of the second Piola-Kirchhoff
stress can be obtained by derivatives of the strain energy density
function,
\begin{eqnarray}
\mathit{S}_{ij}=\frac{\partial\rho_{0}\mathit{W}}{\partial\mathit{E}_{ij}}\,\label{eqn:sixteen}.
\end{eqnarray}
In a simple uniaxial tensile test, the only non-zero component of
stress, denoted by $\mathit{S}$, is
\begin{eqnarray}
\mathit{S}=\mathit{S}_{11}=\frac{\mathit{F}}{\lambda\mathit{A}_{0}}\,,
\end{eqnarray}
where $\mathit{A}_{0}$ and $\mathit{F}$ are the initial
cross-sectional area sample and the load applied to the specimen,
respectively.

Equation (~\ref{eqn:sixteen}) is a constitutive law or
stress-strain relationship. Thus we need to know the details of
the strain-energy functions. For the strain energy function,
studies on non isotropic tissues as skin, muscles, ligaments,
etc.,  have shown that the exponential form applies well. A
general strain-energy function for biological soft tissues is
given by Fung \cite{fung} in the form
%\begin{widetext}
\begin{eqnarray}
\begin{array}{ll}
\mathit{W}&=\frac{1}{2}\alpha_{ijkl}\mathit{E}_{ij}\mathit{E}_{kl}+
\left(\beta_{0}+\beta_{mnpq}\mathit{E}_{mn}\mathit{E}_{pq}\right)\\
&\times\mathit{exp}\left(\gamma_{ij}\mathit{E}_{ij}+\gamma_{ijkl}\mathit{E}_{ij}\mathit{E}_{kl}+\ldots\right)\,,
\label{eqn:eighteen}
\end{array}
\end{eqnarray}
%\end{widetext}
where
$\alpha_{jkl},\,\beta_{0},\,\beta_{mnpq},\,\gamma_{ij},\ldots$ are
constants.

Regarded the complex form of Eqn. (~\ref{eqn:eighteen}) high order
terms can be dismissed as for skin, considered as being in a
bi-axial state, and its strain-energy function is,
\begin{widetext}
\begin{eqnarray}
\begin{array}{ll}
\rho_{0}\mathit{W}&=\frac{1}{2}\left(\alpha_{1}\mathit{E}^{2}_{11}+
\alpha_{2}\mathit{E}^{2}_{22}+\alpha_{3}\mathit{E}^{2}_{12}+
\alpha_{3}\mathit{E}^{2}_{21}+2\alpha_{4}\mathit{E}_{11}\mathit{E}_{22}\right)\\
&\frac{1}{2}c\,exp\left(a_{1}\mathit{E}^{2}_{11}+a_{2}\mathit{E}^{2}_{22}+
a_{3}\mathit{E}^{2}_{12}+a_{3}\mathit{E}^{2}_{21}+2a_{4}\mathit{E}_{11}\mathit{E}_{22}+
\gamma_{1}\mathit{E}^{3}_{11}+\gamma_{2}\mathit{E}^{3}_{22}+
\gamma_{4}\mathit{E}^{2}_{11}\mathit{E}_{22}+\gamma_{5}\mathit{E}_{11}\mathit{E}^{2}_{22}
\right)\,,\label{eqn:nineteen}
\end{array}
\end{eqnarray}
\end{widetext}
where
$\alpha_{1},\ldots,\alpha_{4},a_{1},\ldots\,a_{4},\gamma_{1},\ldots,\gamma_{5}$
and $c$ are constants.

Following, using Eqn. (~\ref{eqn:sixteen}), we obtain the
stresses,
\begin{eqnarray}
\begin{array}{ll}
\mathit{S}_{11}&=\frac{\partial\rho_{0}\mathit{W}}{\partial{E}_{11}}=
\alpha_{1}\mathit{E}_{11}+\alpha_{4}\mathit{E}_{22}+c\mathit{A}_{1}\mathit{X}\,,\\
\mathit{S}_{22}&=\frac{\partial\rho_{0}\mathit{W}}{\partial{E}_{22}}=
\alpha_{4}\mathit{E}_{11}+\alpha_{2}\mathit{E}_{22}+c\mathit{A}_{2}\mathit{X}\,,\\
\mathit{S}_{12}&=\frac{\partial\rho_{0}\mathit{W}}{\partial{E}_{12}}=
\alpha_{3}\mathit{E}_{12}+c\,a_{3}\mathit{E}_{12}\mathit{X}\,,
\end{array}
\end{eqnarray}
where $\mathit{A}_{1},\mathit{A}_{2}$ and $\mathit{X}$ are
functions of the strain components.

It can be verified from Figure (~\ref{fig:three}) that Eqn.
(~\ref{eqn:nineteen}) is suitable for bi-axial mechanical behavior
of rabbit skin. Additional terms can be disregarded taking in
account physical considerations, if we are concerned mainly with
higher stresses and strains the first group of terms in Eqn.
(~\ref{eqn:nineteen}) can be omitted, in addition for practical
purposes third order terms in the exponential function can also be
omitted, and we have simply,
\begin{widetext}
\begin{eqnarray}
\rho_{0}\mathit{W}=\frac{1}{2}c\,exp\left(a_{1}\mathit{E}^{2}_{11}
+a_{2}\mathit{E}^{2}_{22}+a_{3}\mathit{E}^{2}_{12}+a_{3}\mathit{E}^{2}_{21}
+2a_{4}\mathit{E}_{11}\mathit{E}_{22}\right)\,.
\end{eqnarray}
\end{widetext}

\subsection{Viscoelastic Behavior}
\begin{figure}[b]
\centering \vspace*{2.cm}\setlength{\abovecaptionskip}{-3cm} %
\includegraphics[width=8.cm,height=8.cm]{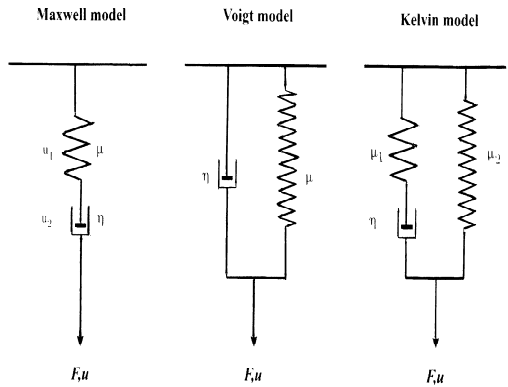}
\caption{Spring-dashpot arranged as to describe different basic
viscoelastic models.} \label{fig:eight}
\end{figure}

Hysteresis, relaxation, and creep are common features of
viscoelastic behavior for many materials. The toy models here
presented are basic mechanical models useful describing
viscoelastic behavior. These are composed by a combination of
elastic material with elastic constant $\mu$ (i.e. spring) and a
dashpot containing a fluid with coefficient of viscosity $\eta$. A
force $F$ is suppose to produce on the spring an uniaxial linear
deformation, denoted by the displacement $\mathit{u}$, and
proportional to $\mu$. Whereas, at any instant, for the dashpot,
it produces a velocity proportional to $\eta$. Different
spring-dashpot arrangements are shown in Figure(~\ref{fig:eight}).

It is remarkable that these classes of discrete models are also
useful in the continuum, where relaxation functions are borrowed
from these theories. For example modelling viscoelastic tissue
behavior is assumed that the stresses in the material, which
themselves may result from a nonlinear stress-strain relation are
linearly superposed with respect to time, recalling the
quasi-linear viscoelastic formulation, which we present next in
subsection (~\ref{subsec:quase-linear visc}).

\subsubsection{Maxwell Model}
In this model an spring and a dashpot are arranged in serie. Thus,
the material behaves like an elastic at short time and it is
viscous at long times.
\begin{eqnarray}
\begin{array}{lr}
\dot{\mathit{u}}=\dot{F}/\mu \,+\,F/\eta \,\,;&\mathit{u}(0)=
\frac{F(0)}{\mu }\,.\label{eqn:twentytwo}
\end{array}
\end{eqnarray}
Equation (~\ref{eqn:twentytwo}) is solved, prescribed the force
$F(t)$ is a unit step function, and the result is called the creep
function $\mathit{c}(t)$. It represents the elongation produced by
a sudden application of a constant unitary force at $t=0$,
\begin{eqnarray}
c(t)=\left( \frac{1}{\mu }+\frac{1}{\eta }t\right)
\mathbf{1}(t)\,.\label{eqn:twentythree}
\end{eqnarray}
Interchanging the role of $F$ and $\mathit{u}$, we obtain the
relaxation function $\mathit{g}(t)$, as a response to a unity
elongation, i.e. $\mathit{u}(t)= \mathbf{1}(t)$. The relaxation
function is the force that must be applied to the specimen in
order to produce an elongation, that changes at $t=0$ from zero to
unity, remaining unity thereafter.
\begin{eqnarray}
\mathit{g}(t)=\mu\,\exp{\left(-\mu{t}/\eta\right)}\,\mathbf{1}(t)\,,\label{eqn:twentyfour}
\end{eqnarray}
with, \textbf{1}(t) is the step function,
\begin{eqnarray}
\mathbf{1}(t)=\left\{
\begin{array}{ccl}
1 & ; & t>0\,, \\
1/2 & ; & t=0\,, \\
0 & ; & t<0\,.
\end{array}
\right.
\end{eqnarray}
In this way, for the Maxwell model we found that a sudden
application of a load induce an immediate deflection of the
elastic spring, followed by a creep of the dashpot, as in Eqn.
(~\ref{eqn:twentythree}). On the other hand, a sudden deformation
produces an immediate reaction by the spring followed by stress
relaxation according to exponential law, which relaxation time is
given by the factor $\eta/\mu$, as in Eqn.
(~\ref{eqn:twentyfour}).

\subsubsection{Voigt Model}
Let us now regard the spring and the dashpot in parallel,
\begin{eqnarray}
\begin{array}{lr}
F=\mu\mathit{u}+\eta\,\dot{\mathit{u}}\,\,;&\,\mathit{u}(0)=0\,.\label{eqn:twentysix}
\end{array}
\end{eqnarray}
Solving Eqn. (~\ref{eqn:twentysix}) we obtain the creep and the
stress relaxations functions, respectively,
\begin{eqnarray}
\begin{array}{ccl}
\mathit{c}(t) & = &
\frac{1}{\mu}\left\{1-\exp{\left(-\mu\,t/\eta\right)}\right\}%
\mathbf{1}(t)\,, \\
\mathit{g}(t) & = & \eta\,\delta(t)+\mu\,\mathbf{1}(t)\,.
\end{array}\label{eqn:twentyseven}
\end{eqnarray}
Where, $\delta(t)$ is the Dirac delta function. Regarding Eqn.
(~\ref{eqn:twentyseven}) we found that, a sudden application of a
unity force will produce no immediate spring deflections; which
solely built up gradually, because the dashpot arranged in
parallel with the spring will not move instantaneously. The
displacement of the dashpot will relax exponentially with a
relaxation time $\eta/\mu$.

\subsubsection{Kelvin Model}
This model is also known as the Standard Linear model. While
models due to Maxwell and Voigt suppose fluids behave like elastic
to some extent, the standard model goes further taking in account
the dissipation rate of energy in various materials subjected to
cyclic loading,
\begin{eqnarray}
\begin{array}{lr}
F\,+\,\tau_{\epsilon}\,\dot{F}\,=\,E_{R}\left(\mathit{u}+\tau_{\sigma}
\dot{\mathit{u}}\right) \,,&\tau_{\epsilon}\,F(0)\,=\,E_{R}\,
\tau_{\sigma}\,\mathit{u}(0)\,.
\end{array}
\end{eqnarray}
Here $\tau_{\epsilon}$ is the time of relaxation of the load under
the condition of constant deflection, whereas $\tau_{\sigma}$ is
the time of relaxation of deflection under the condition of
constant load, and
\begin{eqnarray}
\begin{array}{cc}
\mathit{c}(t) & =\frac{1}{E_{R}}\left\{1-\left(1-\frac{\tau_{\epsilon}}{\tau_{\sigma}}%
\right)e^{-t/\tau_{\sigma}}\right\}\mathbf{1}(t)\,, \\
\mathit{g}(t) & =E_{R}\left\{1-\left(1-\frac{\tau_{\sigma}}{\tau_{\epsilon}}%
\right)e^{-t/\tau_{\epsilon}}\right\}\mathbf{1}(t)\,.
\end{array}\label{eqn:twentynine}
\end{eqnarray}
In Eqn. (~\ref{eqn:twentynine}) as $t\rightarrow\infty$, the
dashpot is completely relaxed and the load-deflection relaxation
becomes that of the springs, characterized by the constant
$E_{R}$, which is called the relaxed elastic modulus.

\subsubsection{Boltzmann Model}
This is a most general formulation under the assumption of
linearity between cause and effect. In the $1-\mathcal{D}$ case,
analogously to the case above, we consider a bar subjected to a
force $F(t)$ and an elongation $\mathit{u}(t)$. The elongation is
caused by the total history of the loading up to the time $t$.
Lets now consider an small time interval $d\tau$; at time $\tau$,
the increment of loading is $\left(\frac{dF}{d\tau}\right)d\tau$ .
It contributes an element $d\mathit{u}(t)$ to the elongation at
time $t$, with a proportionality constant $c$, depending on the
time interval $t-\tau$ , then,
\begin{eqnarray}
\mathit{u}(t)=\int_{0}^{t}\mathit{c}(t-\tau)\,\frac{dF}{d\tau}\,d\tau\,.
\end{eqnarray}
Analogously, for the force,
\begin{eqnarray}
F(t)=\int_{0}^{t}\mathit{g}(t-\tau)\,\frac{d\mathit{u}(\tau)}{d\tau}\,d\tau\,.
\end{eqnarray}
Here the functions $\mathit{c}(t-\tau)$ and $\mathit{g}(t-\tau)$
are the creep function and the relaxation function, respectively.
More generally, we can write the relaxation function in the form,
\begin{eqnarray}
\mathit{g}(t)=\sum_{n=0}^{N}\alpha_{n}\,e^{-t\nu_{n}}\,,\label{eqn:boltzmann}
\end{eqnarray}
where $\nu_{n}=1/\tau_{n}$ is called characteristic frequency and
$\alpha_{n} $ is an amplitude associated to each frequency
$\nu_{n}$, it is called a spectrum of the relaxation function. A
generalization to a continuum spectrum is giving in the next
section.

\subsection{\label{subsec:quase-linear visc}Quasi-Linear Viscoelasticity
of Biological Tissues}
The hysteresis $\mathit{H}$ is due to viscoelasticity, it is
defined as the ratio of the area of the hysteresis loop divided by
the area under the loading curve. For biological tissues,
$\mathit{H}$ is seen to be variable, but its variation with strain
rate is not large, see Section (~\ref{subsec:visco}). Hence,
typical soft tissue behavior shows non-linearity of the
stress-strain relationship and insensitivity of the material to
strain rate.

Models due to Maxwell and Voigt provide a logarithmic type
functional relationship between hysteresis and frequency ($\nu$,
the inverse of the characteristic time), i.e.,
$\mathit{H}(\nu)\propto\,\ln(\nu)$. It is a decreasing curve for
the former, whereas it is a increasing curve for the later. On the
other hand, the Standard model presents a bell-shaped curve of
hysteresis vs the logarithm of the frequency. Summarizing, none of
these models present a typical flat hysteresis vs frequency curve
of living tissues. However, this fact can be corrected by
introducing nonlinear springs. Hence, this now suitable model for
soft tissues, is composed of a long series of Kelvin bodies whose
characteristic times ($1/\nu$) span a broad range.

Lets consider an elastic stress tensor $\mathit{T}^{(e)}$, which
is a function of the strain tensor $\mathit{E}$. If the material
is in the zero-stress state until the time $t=0$, then it is
suddenly strained to $\mathit{E}$ and maintained constant at that
value, thus the stress developed will be a function of time as
well as of $\mathit{E}$. Hence, the history of the stress
development may be written as,
\begin{eqnarray}
\begin{array}{lr}
\mathit{G}_{ijmn}(t)\mathit{T}_{mn}^{(e)}\mathit{(E)}\,;&\,\mathit{G}_{ijmn}(0)=1\,,
\end{array}
\end{eqnarray}
in which $\mathit{G}_{ijmn}(t)$ is called the relaxation function,
a normalized function of time. Now, we suppose that the stress
response to an infinitesimal changes in a component of strain,
$\delta\mathit{E}_{ij}$, superposed on a specimen in a state of
strain $\mathit{E}$ at an instant of time $\tau$, is, for
$t\,>\,\tau \,,$
\begin{eqnarray}
\mathit{G}_{ijmn}(t-\tau )\,\frac{\partial \mathit{T}_{mn}^{(e)}\left[ \mathit{E}%
(\tau )\right]
}{\partial\mathit{E}_{ij}}\partial\mathit{E}_{ij}(\tau )\,.
\end{eqnarray}
In addition, assuming that the stress at time $t$ is the sum of the
contributions of all the past changes,each governed by the same reduced
relaxation function,
\begin{eqnarray}
\mathit{T}_{ij}(t)=\int_{-\infty }^{t}\mathit{G}_{ijmn}(t-\tau
)\,\frac{\partial \mathit{T}_{mn}^{(e)}[\mathit{E}(\tau
)]}{\partial\mathit{E}_{kl}}\frac{\partial\mathit{E}_{kl}(\tau
)}{\partial \tau }\,d\tau \,.
\end{eqnarray}
Although $\mathit{T}^{(e)}(\mathit{E})$ may be a nonlinear
function of strain, the relaxation process is linear. Hence, this
is a quasi-linear viscoelastic theory.

In particular, for the one-dimensional Kelvin model,
\begin{eqnarray}
\mathit{G}(t)=\frac{1+\mathit{S}(q)\exp{\left(-t/q\right)}}{1+\mathit{S}(q)}\,,
\end{eqnarray}
with $q$ and $\mathit{S}(q)$ are constant parameters. The latter
is called the relaxation spectrum and $1/q$ is a frequency. For an
infinite number of Kelvin model in series, we can get the follow
reduced relaxation function in the form,
\begin{eqnarray}
\mathit{G}(t)=\frac{1+\int_{0}^{\infty}%
\mathit{S}(q)\exp{\left(-t/q\right)}\,dq}
{1+\int_{0}^{\infty}\mathit{S}(q)\,dq}\,.
\end{eqnarray}
It has been shown that a specific spectrum, with constants $c,\,q_{1},\,q_{2}
$, the follow function fits the data for the skin,arteries,ureter and teniae
coli.
\begin{eqnarray}
\mathit{S}(q)\left\{
\begin{array}{ccc}
c/q\, & ; & \mbox{$q_{1}\le\,q\le\,q_{2}$}\,, \\
0\, & ; & \mbox{$q<q_{1}\,;q>q_{2}$}\,.
\end{array}
\right.
\end{eqnarray}

\section{\label{sec:vibrsyt}Vibrational Systems}
In this section we suppose that a physical body is conformed by an
ensemble of connected springs. Initially is being considered both
spring elasticity and spring dumping responses are linear.

Hence, let us consider an elastic body obeying Hooke's law. Let it
be rigidly supported in a fixed space and subjected to a set of
forces $\mathit{Q}_{i}$ acting at points
$i\in\left(1,\ldots,n\right)$. Lets set up a generalized
displacement $\mathit{q}_{i}$ at point $i$, in the direction of
the force $\mathit{Q}_{i}$, be
linearly proportional to the forces $\mathit{Q}_{1},\ldots,\mathit{Q}_{n}$, and vice versa, for $%
i\in\left(1,\ldots ,n\right)$:
\begin{equation}
\begin{array}{ccccc}
\mathit{q}_{i}= & \mathit{C}_{i1}\mathit{Q}_{1} & +\mathit{C}_{i2}\mathit{Q}_{2} & +\,...\, & +\mathit{C}_{in}\mathit{Q}_{n}\,, \\
\mathit{Q}_{i}= & \mathit{K}_{i1}\mathit{q}_{1} &
+\mathit{K}_{i2}\mathit{q}_{2} & +\,...\, &
+\mathit{K}_{in}\mathit{q}_{n}\,.
\end{array}
\label{eqn:vib-lin}
\end{equation}
The constants of proportionality $\mathit{C}_{ij},\mathit{K}_{ij}$
are independents of the forces and displacements.
$\mathit{C}_{ij}$ are the flexibility influence coefficients and
$\mathit{K}_{ij}$ are the stiffness influence coefficients,
respectively. The physical meaning of $\mathit{K}_{ij}$ is the
force required to act at the point $i$ due to a unit displacement
at the point $j$, while another points (others than $j$) are held
fixed. Analogously, $\mathit{C}_{ij}$ is the deflection at $i$ due
to a unit force acting at $j$. The constraining equations
(~\ref{eqn:vib-lin}) imply that exist a unique unstressed state to
which the body return whenever all the external forces are
removed, so the superposition principle applies. Moreover, it
implies that the total work done by a set of forces does not
depend of the order in which the forces are applied. This work
done by each force is $\frac{1}{2}Q_{i}q_{i}$ and the total work
done by the system of forces, is stored as strain energy in the
elastic body , and it is given by,
%\begin{widetext}
\begin{eqnarray}
\begin{array}{ll}
\mathcal{U}&=\frac{1}{2}\sum_{i=1}^{n}\mathit{Q}_{i}\mathit{q}_{i}
=\frac{1}{2}\sum_{i=1}^{n}\sum_{j=1}^{n}\mathit{K}_{ij}\mathit{q}_{i}\mathit{q}_{j}\,,\\
&=\frac{1}{2}\sum_{i=1}^{n}\sum_{j=1}^{n}\mathit{C}_{ij}\mathit{Q}_{i}\mathit{Q}_{j}\,.
\end{array}\label{eqn:elastic}
\end{eqnarray}
%\end{widetext}
 The matrices $\mathit{K}_{ij}$ and
$\mathit{C}_{ij}$ are symmetric. In others words, the displacement
at a point $i$ due to a unit force acting at another point $j$ is
equal to the displacement at $j$ due to a unit force acting at
$i$,\,i.e., they are positives in the same direction at each
point.The stability of the system is guarantied if all the
principal minors, including the determinant of the full
$\mathit{K}_{ij}\,\,(\mathit{C}_{ij})$ matrices are positives.

Now we can write the equation of motions of a set o masses
embedded in an elastic body. The D'Alambert's principle
establishes that the particles can be considered to be in a state
of equilibrium if the inertial forces are applied in reversed
direction on the particles. Thus, if the body is attached to a
stationary support and we consider dumpers (lets said, dumping
forces), then the dynamic equations written in vectorial form are
\begin{equation}
m_{i}\ddot{\mathit{q}}_{i}+\sum_{j=1}^{n}\mathit{K}_{ij}\mathit{q}_{j}+\sum_{i=1}^{n}%
\beta_{ij}\dot{\mathit{q}}_{j}\,=\,\mathit{F}_{i}^{e}\,;\left(
i=1,\ldots \,n\right)\,. \label{eqn:mass-spring}
\end{equation}
Here $\mathit{q}_{i}$ is the displacement of the particle and
$\mathit{F}^{e}_{i}$ is the external force acting on it.

However, damping force could be much more complex than this, for
example it can be aerodynamics in origin, nonlinearly viscoelastic
(as for tissues) and not necessarily stabilizing. For an immediate
generalization of the theory,  let recall for the concept of
kinetic energy $\mathcal{K}$, it is an homogeneous quadratic
function of the $\dot{\mathit{q}}$'s, and its rate can be computed
from,
\begin{eqnarray}
\dot{\mathcal{K}}=\sum_{i=1}^{n}\left( \frac{d}{dt}\frac{\partial \mathcal{K}}{%
\partial {\dot{\mathit{q}}}_{i}}-\frac{\partial \mathcal{K}}{\partial
\mathit{q}_{i}}\right) \dot{q}_{i}\,.
\end{eqnarray}

For a biological system, other forms of energy rather than the kinetic
energy $\mathcal{K}$ can also be involved, such as the gravitational potential $%
\mathcal{G}$, the internal energy $\mathcal{U}$, and the chemical energy $%
\mathcal{C}$. According to the first law of thermodynamics, the
energy of a system solely can be changed by absorption of heat
$\mathcal{H}$ and by doing work on the system. The rate of change
of the total energy must be equal to the sum of the rates of heat
input $\dot{\mathcal{H}}$ and work done on the system
$\dot{\mathcal{W}}$, so,
\begin{eqnarray}
\dot{\mathcal{K}}\,+\,\dot{\mathcal{G}}\,+\,\dot{\mathcal{U}}\,+\,\dot{%
\mathcal{C}}\,=\,\dot{\mathcal{H}}\,+\,\dot{\mathcal{W}}\,.\label{eqn:balance}
\end{eqnarray}
Generally, just a certain part of the internal energy function
$\mathcal{U}$ arise from the strain energy, it is a quadratic
function of (generalized) coordinates $q_{i} $. It can also depend
on the temperature. If the temperature remains constant, then the
rate at which work is done by the internal energy and also by the
forces acting on a system are, respectively,
\begin{eqnarray}
\begin{array}{lr}
\dot{\mathcal{U}}\,=\,\sum_{i=1}^{n}\frac{\partial\mathcal{U}}{\partial \,q_{i}}\,%
\dot{q}_{i}\,;&\dot{\mathcal{W}}\,=\,\sum_{i=1}^{n}Q_{i}\,\dot{q}%
_{i}\,,
\end{array}
\end{eqnarray}
then, substituting them into Eqn. (~\ref{eqn:balance}),
\begin{equation}
\sum_{i=1}^{n}\left( \frac{d}{dt}\frac{\partial \mathcal{K}}{\partial \dot{q}%
_{i}}-\frac{\partial \mathcal{K}}{\partial \,q_{i}}+\frac{\partial \mathcal{U%
}}{\partial \,q_{i}}-Q_{i}\right) \dot{q}_{i}=\dot{\mathcal{H}}-\dot{%
\mathcal{G}}-\dot{\mathcal{C}}\,.  \label{eqn:eq-geral00}
\end{equation}
Now lets consider the special case when $\dot{\mathcal{H}}-\dot{\mathcal{G}}-%
\dot{\mathcal{C}}\equiv \,0\,.$ Because the displacements are independent
variables, that can assume arbitrary variables, we impose for any $\dot{q}%
_{i}\ne \,0$, in Eqn. (~\ref{eqn:eq-geral00}),
\begin{eqnarray}
\frac{d}{dt}\frac{\partial \mathcal{K}}{\partial \dot{q}_{i}}\,-\,\frac{%
\partial \mathcal{K}}{\partial \,q_{i}}\,+\,\frac{\partial \mathcal{U}}{%
\partial \mathit{q}_{i}}\,=\,Q_{i}\,.\label{eqn:eq-geral01}
\end{eqnarray}
If the (generalized) forces are partial derivatives of a function, i.e., $%
Q_{i}=-\frac{\partial \mathcal{W}}{\partial \,q_{i}}$; where
$\mathcal{W}$ is the potential energy. For the spring body system
the internal elastic energy is given by Equation
(~\ref{eqn:elastic}).
Thought, $\mathcal{W}$ and $\mathcal{U}$ only depend on $%
\mathit{q}_{i}$, the system is conservative.

The aerodynamic or hydrodynamics forces acting on an animal moving
in a fluid are not conservative. Hence, for problems involving
these forces is much better to use Equation
(~\ref{eqn:eq-geral01}), thought $\mathit{Q}_{i}$ is an external
force.

Following, we apply the present theory to materials showing
viscoelastic behavior (like tissues), in the presence of fluid
dynamic forces.

\subsection{\label{sec:vibsyst-fuid}Systems with Dumping and Fluid Dynamics Loads}
Here we recall the most general constitutive equation of a linear
viscoelastic solid reviewed in Section~\ref{subsec:visco}.
Applying Boltzmann's formulation we may write the internal force
acting on a particle $i$, located at point $i$, by a displacement
$\mathit{q}_{j}$ located at point $j$ as,
\begin{eqnarray}
\mathit{F}_{ij}(t)=\int_{-\infty}^{t}\,G_{ij}(t-\tau)\frac{d\mathit{q}_{j}(\tau)}{%
d\tau}d\tau\,,\label{eqn:viscocity}
\end{eqnarray}
where $\mathit{G}_{ij}(t)$ are the relaxation functions. For solid
with fading memory, it is given by equation
(~\ref{eqn:boltzmann}),
\begin{eqnarray}
\mathit{G}_{ij}(t)=\mathit{K}_{ij}+\beta_{ij}e^{-t\nu_{1}}+\,...\,+\gamma_{ij}e^{-t\nu_{n}}\,,
\end{eqnarray}
where $\nu_{1},\nu_{2},\,...\,,\nu_{n}$ are the relaxation frequencies , $%
K_{ij},\beta_{ij},\gamma_{ij}$ are spectral constants.

For systems subjected to fluid dynamics forces (i.e., in swimming, in wind
or with internal flow), the external force at the point $i$ due to motion at
point $j$ may be written as,
\begin{eqnarray}
F_{ij}^{(e)}(t)=\int_{-\infty}^{t}\,A_{ij}(t-\tau)\frac{d\mathit{q}_{j}(\tau)%
}{d\tau}d\tau\,,\label{eqn:fluid}
\end{eqnarray}
where $A_{ij}(t)$ are aerodynamic influence functions, which is un-symmetric.

Considering both the internal forces and a fluid dynamic forces
given by Eqn.(~\ref{eqn:viscocity}) and Eqn. (~\ref{eqn:fluid}),
Eqn. (~\ref{eqn:eq-geral01}) can be rewritten as,
%\begin{widetext}
\begin{eqnarray}
\begin{array}{ll}
&m_{i}\frac{d^{2}\textit{q}_{i}(t)}{dt^{2}}+\sum_{j=1}^{n}\int_{-\infty}^{t}\mathit{G}_{ij}(t-\tau)\frac{d\textit{q}_{j}(\tau)}{d\tau}d\tau\\
&=\sum_{j=1}^{n}\int_{-\infty}^{t}\,A_{ij}(t-\tau)\frac{d\textit{q}_{j}(\tau)}{d\tau}d\tau+F_{i}^{(e)}
\end{array}
\end{eqnarray}
%\end{widetext}
where $F^{(e)}_{i}$ is the external force other than fluid dynamic forces.

\section{\label{sec:conclus}Conclusions and Perspectives}
More general mechanical properties of a material can be specified
by mean of constitutive equations. Remarkable, the continuum
approach naturally adds constitutive equations into more
generalized theories. However, it seems that soft tissue models
based on these more general theories are numerically expensive,
hence that linear elasticity approaches have been preferentially
considered.

On the other hand, spring-mass models present another alternative
to model soft tissue mechanical behavior. Because of physical
model simplicity and easier implementation, it has been
preferentially used on applications for surgical simulations. The
method is economic and fast, so it is a good candidate to operate
in real time. However, until it is of our knowledge, its numerical
implementation has been done no far away from the limits of
linearity, not only regarding elasticity but even viscoelasticty.

In this work we have review some possible alternative models
considering nonlinear elasticity and quasi-linear viscoelasticity
on spring-mass models. The former can be naturally filling up by
adding to the model exponential type deformations, as in section
~\ref{tissuemath-spring}. Viscoelastic mechanical behavior can be
added to the model using quasi linear models based on Boltzmann
type approach, as shown in section ~\ref{sec:vibsyst-fuid}.
Summarizing, these are the formulations we are going to explore in
order to better model soft tissues. The resulting model will be
validate by comparison with accurate models based on the continuum
approach. We shall stress that because of simplicity and economy,
this more elaborated spring-mass models can still handle to
perform real time calculations.

Recently, Holzapfeld \cite{holzapfel3} and Ogden \cite{ogden2}
have investigated numerical method based on the Continuum approach
in order to regard nonlinear elasticity and Viscoelastic response
on arterial walls and soft tissues. In addition to research for
more sophisticated spring-mass models, these numerical methods are
also going to pave the way of our research.

\begin{acknowledgments}
 We thanks CNPq for support.
\end{acknowledgments}


\begin{thebibliography}{99}
\bibitem{ogden}Ogden, R.W. (2001) Nonlinear Elasticity, Anisotropy,
Material Stability and Residual Stress in Soft Tissue.

\bibitem{holzapfel}Holzapfel, G.A. (2000) Nonlinear Solid Mechanics.
Chichester: Wiley.

\bibitem{holzapfel2}Holzapfel, G. A. (2001) Biomechanics of Soft Tissue.
In Lecture, J.ed., Handbook of Material Behavior Models. Boston: Academic
Press 1049-63.

\bibitem{bro-nielsen}Bro-Nielsen, M. (1998). Finite Element Modelling in
Surgery Simulation. Proc. of the IEEE: Special Issue on Virtual and
Augmented Reality in Medicine 86 (3) 524-30.

\bibitem{keeve}Keeve, E., Girod, S., Pfeifel, P., Girod, B. (1996)
Anatomy-Based Facial Tissue Modelling using the Finite Element Method. Proc.
IEEE Visualization, San Francisco, CA. USA.

\bibitem{koch}Koch, R.M., Gross, M.H., Bueren, D.F.,Franhauser, G.,
Parish, Y., Carls, F.R.(1996) Simulation Facial Surgery using Finite Element
Modles. SIGGRAPH'96, ACM Computer Graphycs, August 30 .

\bibitem{mosegaard}Mosegaard, J. (2003). LR-Spring-mass Model for Cardiac
Surgical Simulation. Medicine Meets Virtual Reality, (12) 256-58.

\bibitem{tescher}Teschrer, M., Girod, S., Girod, B. (2000). Direct
Computation of Nolinear Soft-Tissue Deformation. Vison Modeling and
Visualization VMV'000, Saarbucken, Germany, November 22-24.

\bibitem{fung}Fung, Y.C. (1994) A first course in continuum mechanics.
Prentice Hall, Englewood Cliff, New Yersey.

\bibitem{hasegawa}Hasegawa, M, and Azuna, T. (1974). Wall structure and
static viscoelasticities of large veins. J. Jap. College Angiol. 14:87-92.

\bibitem{tong}Tong, P. and Fung, Y.C. (1976). The stress-strain
relationship for the skin. J. Biomech. 9:649-57.

\bibitem{yamamoto}Yamamoto, E., Hayashi, K. and Yamamoto, N. (1999).
Mechanical properties of collagen fascicles from the rabit pattelar tendom.
ASME J. Biomech. Eng. 121:124-31.

\bibitem{chuong}Chuong,C.J., and Fung,Y.C.(1984). Compressibility and
Constitutive equation of arterial wall in radial compression experiments. J.
Biomech, 17:35-40.

\bibitem{woo}Woo, S.L.-Y., Lubock, P., Gomez, M.A.,Jemmott, G.F., Kuei, S.C. and
Akeson, W.H.(1979) Large deformation nonhomogeneus and directional
propertiesof articular cartilage. J. Biomech. 12:437-46.

\bibitem{fung2}Fung,Y.C.,Fronek,K. and Patitucci,P. (1979).
Pseudoelasticity of arteries and the choice of its mathematical expression.
Am. J. Physiol. 237:H620-631.

\bibitem{holzapfel3}Holzapfel, G.A.,(2003). Structural and Numerical
models for the (Visco)elastic Response of Arterial Walls with Residual
Stresses, 109-184.International Centre for Mechanical Sciences. Springer,
Wien, New Yersey.

\bibitem{ogden2}Ogden, R.W.,(2003). Nolinear Elasticity, Anisotropy,
Matyerial Stability and Residual Stress in Soft Tissue.Biomechanical of Soft
Tussue in Cardiovascular Systems, 65-108. International Centre for
Mechanical Sciences. Springer, Wien, New Yersey.
\end{thebibliography}
\end{document}